\documentclass[aps,prl,reprint,superscriptaddress,floatfix]{revtex4-1}

\usepackage{amssymb}
\usepackage{amsmath}
\usepackage{psfrag}
\usepackage{topcapt}
\usepackage{hyperref}

\usepackage[a4paper,hmargin={1.5cm,1.5cm},vmargin={1.5cm,2.5cm}]{geometry} 

\usepackage{ifpdf}

\ifpdf
\usepackage[pdftex]{graphicx}
\else
\usepackage{graphicx}
\fi

\begin{document}

\title{Observation of the Bloch-Siegert Shift in a Qubit-Oscillator System\\ in the Ultrastrong Coupling Regime}

\author{P. Forn-D\'iaz}
\email{p.forndiaz@tudelft.nl}
\affiliation{Quantum Transport Group, Delft University of Technology, Lorentzweg 1, 2628CJ Delft, The Netherlands}
\author{J. Lisenfeld}
\affiliation{Quantum Transport Group, Delft University of Technology, Lorentzweg 1, 2628CJ Delft, The Netherlands}
\affiliation{Physikalisches Institut and DFG Center for Functional Nanostructures (CFN),
Karlsruhe Institute of Technology, Karlsruhe, Germany}
\author{D. Marcos}
\affiliation{Theory and Simulation of Materials, Instituto de Ciencia de Materiales de Madrid, CSIC, Cantoblanco 28049, Madrid, Spain} 
\author{J. J. Garc\'ia-Ripoll}
\affiliation{Instituto de F\'{\i}sica Fundamental, CSIC, Serrano 113-bis, 28006 Madrid, Spain}
\author{E. Solano}
\affiliation{Departamento de Qu\'{\i}mica F\'{\i}sica, Universidad del Pa\'{\i}s Vasco - Euskal Herriko Unibertsitatea, Apdo.\ 644, 48080 Bilbao, Spain}
\affiliation{IKERBASQUE, Basque Foundation for Science, Alameda Urquijo 36, 48011 Bilbao, Spain}
\author{C. J. P. M. Harmans} 
\affiliation{Quantum Transport Group, Delft University of Technology, Lorentzweg 1, 2628CJ Delft, The Netherlands}
\author{J. E. Mooij}
\affiliation{Quantum Transport Group, Delft University of Technology, Lorentzweg 1, 2628CJ Delft, The Netherlands}

\date{\today}

\begin{abstract}
We measure the dispersive energy-level shift of an $LC$ resonator magnetically coupled to a superconducting qubit, which clearly shows that our system operates in the ultrastrong coupling regime. The large mutual kinetic inductance provides a coupling energy of $\approx0.82$~GHz, requiring the addition of counter-rotating-wave terms in the description of the Jaynes-Cummings model. We find a 50~MHz Bloch-Siegert shift when the qubit is in its symmetry point, fully consistent with our analytical model. 
\end{abstract}

\maketitle

The study of driven two-level systems has been at the heart of important discoveries of fundamental effects, both classical and quantum mechanical. 
A generic example is the field of nuclear magnetic resonance where the dynamics of nuclear spins is controlled by the application of radio frequency pulses, 
resulting in coherent Rabi oscillations of the spin moments \cite{cohen_book}. In the usual description, the applied harmonic field is decomposed into two mutually counterrotating fields. At resonance in the weak-driving limit only the corotating component interacts constructively with the spins, leading to a Rabi frequency that scales linearly with the driving strength. Thus for this single component corotating regime the rotating-wave approximation (RWA) is known to hold. If the driving is so strong that the Rabi frequency approaches the Larmor frequency, the counterrotating terms need to be taken into account. This leads to an energy shift in the level transition, denoted as the Bloch-Siegert shift \cite{bloch, klimov}. This non-RWA regime has been observed in a variety of strongly driven systems. In the field of quantum electrodynamics (QED) a quantum Bloch-Siegert shift has been considered for atoms very strongly coupled to single photons 
\cite{shirley}, although the experimental verification is difficult \cite{zela}. In the dispersive regime this shift is sometimes referred to as dynamical Stark 
shift \cite{klimov}. For an atom that resides in a resonant cavity the interaction strength $g$, the Rabi rate when the cavity contains a single photon, is found to be typically $10^{-4}$ of the atomic Larmor frequency $\omega_q/2\pi$ and the cavity frequency $\omega_r/2\pi$. The Jaynes Cummings (JC) model \cite{JC}, that fully relies on the validity of the rotating-wave approximation, therefore yields a good description of the system \cite{cqed}. 

In circuit QED \cite{blais} superconducting qubits play the role of artificial atoms. With energy level transitions in the microwave regime, they can be easily cooled to the ground state at standard cryogenic temperatures. These ``atoms'' can interact very strongly with on-chip resonant circuits and reproduce many of the physical phenomena that had been previously observed in cavities with natural atoms \cite{irinel, *johansson, baur, *max}. The large dipolar coupling achievable in superconducting circuits enabled exploring the strong-dispersive limit \cite{schuster}. One now starts addressing the ultrastrong coupling regime $g/\omega_r\sim1$ 
\cite{sahel, borja, garching}. In this Letter we experimentally resolve the quantum Bloch-Siegert shift in an $LC$ resonator coupled to a flux qubit with a coupling strength $g/\omega_r\simeq0.1$, thus entering the ultrastrong coupling regime. 

\begin{figure}[!hbt]
\includegraphics{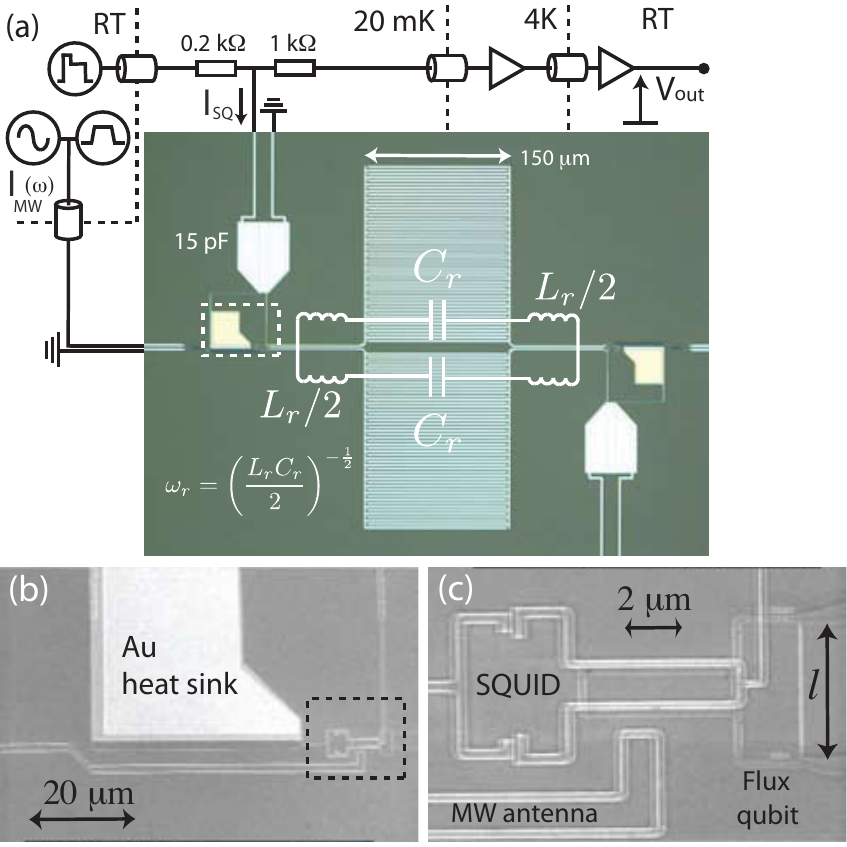}
\caption{\label{fig1}(Color online) Circuit layout and images of the device. (a)~Schematic of the measurement setup. 
The interdigitated capacitor of the $LC$ resonator can be seen in the center of the optical image, with the circuitry of the two SQUIDs next to it (top left and bottom right); $C_r/2\simeq0.25$~pF and $L_r\simeq1.5$~nH. (b)~Scanning electron micrograph (SEM) picture of the SQUID circuit. The readout line is made to overlap with a big volume of AuPd and Au to thermalize the quasiparticles when the SQUID switches. (c) SEM picture of the qubit with the SQUID on top. On the right of the picture the coupling wire to the resonator of length $l$ can be seen.}
\end{figure}

Our system consists of a four-Josephson-junction flux qubit 
\cite{hans}, in which one junction is made smaller than the other three by a factor of approximately 0.5. The qubit is galvanically connected to a lumped-element $LC$ resonator [Fig.~\ref{fig1}]. In previous work the employed $LC$ resonators were strongly coupled to the flux qubit \cite{irinel,*johansson, arkady}, but since they were loaded by the impedance of the external circuit their quality factor was low. Flux qubits have also been successfully coupled to high-quality transmission line resonators \cite{nec,*ibm}. In our experiment we use an interdigitated finger capacitor in series with a long superconducting wire, following the ideas from lumped-element kinetic inductance detectors \cite{simon}. In order to read out the qubit state a dc-switching SQUID magnetometer was placed on top of the qubit. The detection procedure can be found in~\cite{patricedelft}. 

The qubit and the resonator were fabricated in the same layer of evaporated aluminum using standard lithography techniques \cite{patricedelft}. A second aluminum layer galvanically isolated from the first one contains the SQUID and its circuitry together with the microwave antenna to control the local frustration and to produce flux and microwave pulses in the qubit [Fig.~\ref{fig1}]. An external coil is used to generate a magnetic field in the qubit and SQUID in order to bias them at their operating points. A second qubit with its own circuitry was also coupled to the resonator [Fig.~\ref{fig1}~(a)], but during the experiment it was always flux biased such that it did not affect the measurements.

The resonator is made of two capacitors, each containing 50 fingers of 150~$\mu$m length and 1.5~$\mu$m width, separated by 2~$\mu$m [Fig.~\ref{fig1}~(a)]. The two capacitors are linked by two 500~$\mu$m long superconducting wires of 1~$\mu$m width. With these parameters we estimate a capacitance of $C_r\simeq0.5$~pF and an inductance of $L_{r}\simeq1.5$~nH, corresponding to a resonance frequency $\omega_r/2\pi=1/[2\pi\sqrt{L_r(C_r/2)}]\simeq8.2$~GHz. At temperatures $\sim30$~mK the resonator will be mostly in its ground state, with zero-point current fluctuations $\displaystyle I_{\text{rms}}=\sqrt{\hbar\omega_r/2L_r}\simeq40$~nA.

The flux qubit, with an externally applied magnetic flux of $\Phi\approx\Phi_0/2$, behaves effectively as a two-level system ($\Phi_0=h/2e$ is the flux quantum). Since the second excited state is at a much higher energy (typically 30~GHz), the effective Hamiltonian can be written as $H_q=-(\epsilon\sigma_z+\Delta\sigma_x)/2$ using the Pauli matrix notation in the basis of the persistent current states $\lbrace|\circlearrowright\,\rangle$, $|\circlearrowleft\,\rangle\rbrace$. Here $\epsilon=2I_p(\Phi-\Phi_0/2)$, with $I_p$ the persistent current in the qubit loop. $\Delta$ is the tunnel coupling between the two persistent current states. The qubit is inductively coupled to a dc-SQUID detector with a mutual inductance of $M_{\text{SQ}}\simeq5$~pH.

The qubit is galvanically attached to the resonator [Fig.~\ref{fig1}~(c)] with a coupling wire of length $l=5~\mu$m, width $w=100$~nm and thickness $t=50$~nm. To achieve our coupling energy we use the kinetic inductance $L_K$ of the wire that can easily be made larger than the geometric contribution. The kinetic inductance for our narrow dirty wire is found from its normal state resistance \cite{tinkham} $L_K=0.14\hbar R_n/k_BT_c\simeq(25\pm2)$~pH. The strength of the coupling can be approximated by $\hbar g=I_pI_{\mathrm{rms}}L_K$ \cite{lindstrom, bourassa}. Since our $\sim500$ $\mu$m $LC$ resonator is much smaller than the wavelength at the resonance frequency ($\lambda_r\approx20$ mm), the current is uniform along the superconducting wires connecting the capacitor plates. Therefore the position of the qubit along the inductor will not affect the coupling strength. 

The interaction between qubit and resonator can be described by a coupling of dipolar nature $H_{\text{int}} = \hbar g (a^{\dagger} + a) \sigma_z$ in the basis of the persistent current states, where $a^{\dagger}$ ($a$) is the photon creation (annihilation) operator in the basis $\lbrace |n\rangle \rbrace$ of Fock states of the resonator. In the basis of the eigenstates of the qubit, $\lbrace|g\rangle,|e\rangle\rbrace$, the Hamiltonian reads

\begin{multline}
\label{eq1}
H_E=\frac{\hbar\omega_q}{2}\sigma_z+\hbar\omega_r\left(a^{\dag}a+\frac{1}{2}\right)\\+\hbar g\left(\cos(\theta)\sigma_z-\sin(\theta)\sigma_x\right)(a+a^{\dag}),
\end{multline}
with $\hbar\omega_q\equiv\sqrt{\epsilon^2+\Delta^2}$ and $\tan(\theta)\equiv\Delta/\epsilon$. Note that the RWA has not been employed to obtain Eq.~\ref{eq1}.

The qubit-resonator interaction can be rewritten in terms of rising and lowering operators $\sigma_{\pm} = (\sigma_x \pm i \sigma_y)/2$. This yields corotating terms $\sim (\sigma_{+}a + \sigma_{-}a^{\dagger})$ as well as counterrotating terms $\sim (\sigma_{+}a^{\dagger} + \sigma_{-}a)$. In the regime where $g$ is comparable to $\omega_q$ and $\omega_r$, the usual RWA is not valid and the counter-rotating terms cannot be neglected. In order to evaluate their effect on the system, we perform a unitary transformation $H_E'=e^S H e^{-S}$, with $S=\gamma (\sigma_{+} a^{\dagger} - \sigma_{-} a)$ and $\gamma=-g\sin(\theta)/(\omega_q+\omega_r)$ to eliminate the counter-rotating terms. If $|\gamma|\ll1$ we can safely neglect off-resonant terms of order $\gamma^2$. Off-diagonal two-photon processes 
can be removed by similar canonical transformations \cite{klimov}. For frequencies not too far from resonance, keeping terms up to second order in $\gamma$, we obtain the effective Hamiltonian

\begin{figure}[!hbt]
\includegraphics{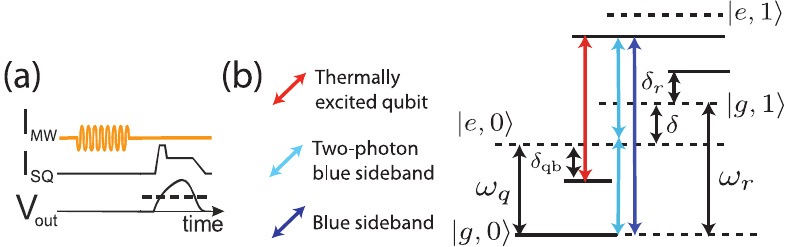}
\caption{\label{fig2}(Color online) Measurement scheme and energy-level diagram. (a)~Schematic of the measurement protocol to perform qubit spectroscopy. (b)~JC ladder depicting the energy-level structure of the system of a flux qubit coupled to an $LC$ resonator. The levels are drawn for the case 
$\delta=\omega_q-\omega_r<0$. The arrows represent the level-transitions that are visible in the spectrum. The dashed lines represent the uncoupled qubit and resonator states. $\delta_q$ and $\delta_r$ are the dispersive shifts that the qubit and the resonator induce to each other.}
\end{figure}

\begin{multline}
\label{eq2}
H_E'=\frac{\hbar\omega_q}{2}\sigma_z+\hbar\omega_r\left(\hat n+\frac{1}{2}\right)+\hbar\omega_{\text{BS}}\left[\sigma_z\left(\hat n+\frac{1}{2}\right)-\frac{1}{2}\right]\\+\hbar g(\hat n)a^{\dag}\sigma_-+\hbar a\sigma_+g(\hat n),
\end{multline}
with $\hat n=a^{\dag}a$. Here the term proportional to $\omega_{BS} \equiv g^2 \sin^2(\theta)/(\omega_q+\omega_r)$ describes the Bloch-Siegert shift. The term $g\cos(\theta)\sigma_z(a+a^{\dag})$ from Eq.~\ref{eq1} has been neglected as to second order it only adds a global phase. The coupling constant has been renormalized to $g(\hat n) \equiv -g\sin(\theta) [ 1 - \hat{n} \; \omega_{BS}/(\omega_q+\omega_r)]$. 

In the basis $\lbrace |g,n+1\rangle, |e,n\rangle\rbrace$, the effective Hamiltonian [Eq.~\ref{eq2}] is box-diagonal. The box corresponding to $n$ photons has eigenvalues
$\lambda_{n,m=0,1} = \hbar\omega_r\left(n+1\right)+(\hbar/2)(-1)^m\sqrt{\delta_{n+1}^2+4g_{n+1}^2}$, 
$\lambda_{0,g} = -\hbar\delta/2$,
where $\delta=\omega_q-\omega_r$ is the detuning, and $\delta_{n+1}=\delta+2\omega_{\text{BS}}(n+1)$ and $g_{n+1}=g\sin(\theta)\sqrt{n+1}[1-(n+1)\omega_{\text{BS}}/(\omega_q+\omega_r)]$. $m=0$ (1) corresponds to the qubit in the ground (excited) state. In the limit $\omega_{\text{BS}}\to0$, the JC result is recovered \cite{blais}. For the qubit in the ground state the oscillator resonance is shifted with respect to the JC model.

\begin{figure}[!hbt]
\includegraphics{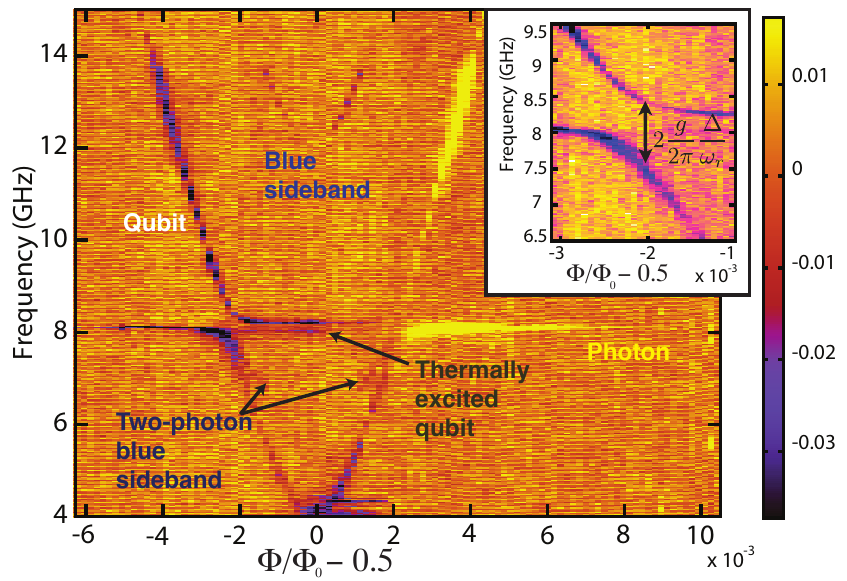}
\caption{\label{fig3}(Color online) Spectrum of the flux qubit coupled to the $LC$ resonator. An avoided-level crossing is observed at a frequency of 8.13~GHz. 
The weak transition near 8~GHz is associated with excited photons due to thermal population of the qubit excited state ($T_{\rm eff}\sim100~$mK at $\sim4-5~$GHz energy splitting). (Inset)~Zoom in around the resonance between qubit and oscillator. The splitting on resonance is $2g\sin(\theta)/2\pi\simeq0.9$~GHz.} 
\end{figure}

\begin{figure}
\includegraphics{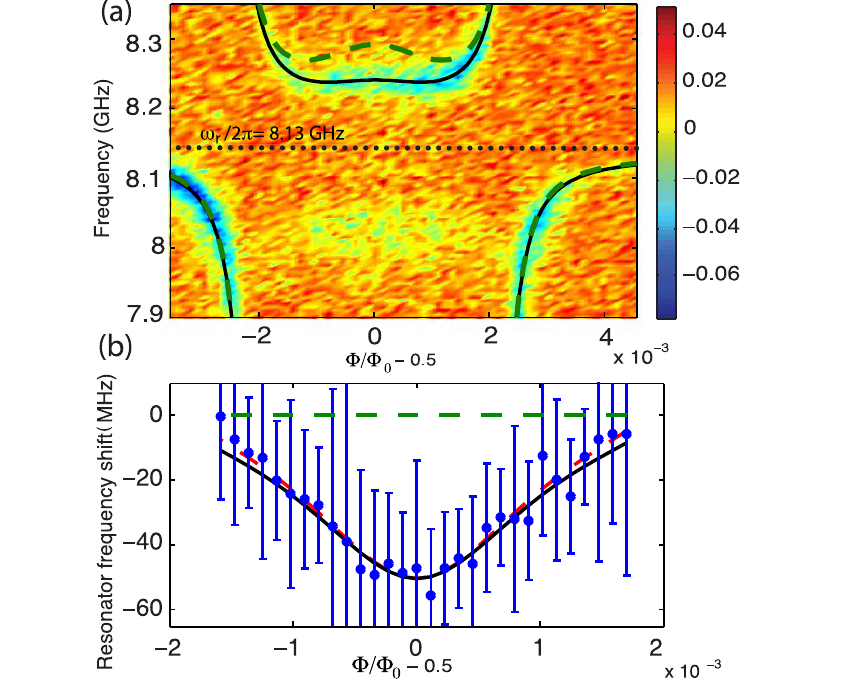}
\caption{\label{fig4}(Color online) Bloch-Siegert shift. (a)~Spectrum in proximity to the resonator frequency obtained using lower driving power than in Fig.~\ref{fig3} and flux pulses \cite{arkady}. The solid black line is the fit of Eq.~\ref{eq1} and the dashed green line is a plot of the JC model (Eq.~\ref{eq1} without counter-rotating terms). The dotted line indicates the bare resonator frequency $\omega_r$. A clear deviation between the dashed line and the data can be observed around the symmetry point of the qubit. A transition associated with thermal population of the qubit excited state can be observed around 8~GHz. (b)~Difference between measurement (blue dots) and the prediction of the JC model (dashed green line). The solid black curve is the same as the solid black curve in (a) and the dashed red curve represents $\lambda_{1,g}-\lambda_{0,g}$. All the curves are subtracted from the JC model. The blue dots are peak values extracted from Lorentzian fits to frequency scans at fixed flux, with the error bars representing the full width at half maximum of each Lorentzian.}
\end{figure}

We prepare the qubit in the ground state by cooling it to 20~mK in a dilution refrigerator. Using the protocol shown in Fig.~\ref{fig2}~(a), we measure the spectrum of the qubit-resonator system [Fig.~\ref{fig3}]. To obtain a higher resolution in the relevant region around 8.15~GHz, we repeated the spectroscopy using lower driving power in combination with the application of flux pulses in order to equalize the qubit signal by reading out far from its degeneracy point [Fig.~\ref{fig4}] \cite{arkady}. We can identify the energy-level transitions on the basis of the JC ladder shown in Fig.~\ref{fig2}~(b). A large avoided crossing between states $|g,1\rangle$ and $|e,0\rangle$ is observed around a frequency of $\sim8$~GHz. This is very close to the estimated resonance frequency of the oscillator. The energy splitting $(2g/2\pi)(\Delta/\omega_r)$ [Fig.~\ref{fig3}~(inset)] is approximately 0.9~GHz. A combined least-squares fit of the full Hamiltonian [Eq.~\ref{eq1}] of the data from Figs.~\ref{fig3}, \ref{fig4} leads to $\Delta/h=(4.20\pm0.02)$~GHz, $I_p=(500\pm10)$~nA, 
$\omega_r/2\pi=(8.13\pm0.01)$~GHz 
and $g/2\pi=(0.82\pm0.03)$~GHz.

The value of $g$ obtained is in good agreement with $I_pI_{\text{rms}}L_K/h=(0.83\pm0.08)$~GHz. Thus 
we find $g/\omega_r\approx0.1$. This large value brings us into the ultrastrong coupling regime, and below we will demonstrate that the system indeed shows ultrastrong coupling characteristics.

The spectral line of the resonator can be resolved when it is detuned several GHz away from the qubit [Fig.~\ref{fig3}]. This could be caused by the external driving when it is resonant with the oscillator. By loading photons in it, the oscillator can drive the qubit off-resonantly by their large coupling. Another possibility is an adiabatic shift during state readout through the anticrossing of the qubit and resonator energies. The qubit readout pulse produces a negative shift of -2 m$\Phi_0$ in magnetic flux, making the spectral amplitude asymmetric with respect to the qubit symmetry point [Fig.~\ref{fig3}]. For our parameters, this shift is coincidental with the avoided level crossing with the oscillator. Then, a state containing one photon in the resonator ({\it{e.~g.}}, $\Phi/\Phi_0-0.5=4$~m$\Phi_0$ in Fig.~\ref{fig3}) can be converted into an excited state of the qubit with very high probability, as the Landau-Zener tunneling rate is very low. Both effects, off-resonant driving and adiabatic shifting, would explain that the sign of the spectral line of the resonator coincides with the one of the qubit on both sides of the symmetry point. Irrespective of the mechanism, the spectral features of Fig.~\ref{fig3} allow us to give a low bound for the quality factor of the resonator $Q>10^3$.

In Figs.~\ref{fig4}~(a),~(b) a marked difference in the resonator frequency between the fit of Eq.~\ref{eq1} (solid black line) and the JC model, Eq.~\ref{eq1} with the counterrotating terms removed, (dashed green line) can be clearly resolved \cite{epaps}. The difference is largest (50~MHz) at the symmetry point of the qubit. This is the Bloch-Siegert shift $\omega_{BS}$ associated with the counterrotating terms [Eq.~\ref{eq2}]. The maximum difference occurs at the symmetry point as outside of it the effective coupling $g\sin(\theta)$ decreases with increasing $\epsilon$. Figure~\ref{fig4}~(b) shows in a dashed red curve a plot of $\lambda_{1,g}-\lambda_{0,g}$ subtracted from the JC model. The agreement between the measured spectral peaks of the resonator and the calculated values using $\lambda_{1,g}-\lambda_{0,g}$ is very good. Concerning the qubit, according to $\lambda_{0,e}-\lambda_{0,g}$ it should experience the same shift $\omega_{BS}$ as the resonator, but with opposite sign. Since the qubit line width at the symmetry point around 4~GHz is very large ($\approx80$~MHz), the Bloch-Siegert shift cannot be clearly resolved there. 

In conclusion, we have measured the Bloch-Siegert shift in an $LC$ resonator strongly coupled to a flux qubit. This demonstrates the failure of the rotating-wave approximation in this ultrastrong coupling regime of circuit QED. The large coupling of 0.82~GHz is achieved using the kinetic inductance of the wire that is shared by the two systems. The coupling could easily be further enhanced by increasing the kinetic inductance or by inclusion of a Josephson junction \cite{garching, bourassa}. This will allow the exploration of the system deeply into the ultrastrong coupling regime where $g$ is comparable with $\omega_r$.

The authors would like to thank R. Aguado for his contributions to the theory, A. Fedorov for fruitful discussions, and R. van Ooijik and R. N. Schouten for assistance in the measurement electronics. We acknowledge financial support from the Dutch NanoNed program, the Dutch Organization for Fundamental Research (FOM), the EU projects EuroSQIP, CORNER and SOLID, the MICINN Projects No. FIS2009-10061 and No. FIS2009-12773, the CAM project QUITEMAD and the Spanish Grants No. FPU AP2005-0720 and No. FIS2009-08744.

\newpage

\onecolumngrid
\begin{center}
{\Large\textbf{Supplementary information: \\Fitting with the Jaynes-Cummings model beyond the rotating-wave approximation}}

\vspace{15pt}
{P. Forn-D\'iaz,$^1$ J. Lisenfeld,$^{1,2}$ D. Marcos,$^3$ J. J. Garc\'ia-Ripoll,$^4$ E. Solano,$^{5,6}$ C. J. P. M. Harmans,$^1$ and J. E. Mooij$^1$}
\vspace{5pt}

\small{\textit{$^1$Quantum Transport Group, Delft University of Technology, Lorentzweg 1, 2628CJ Delft, The Netherlands\\
$^2$Physikalisches Institut and DFG Center for Functional Nanostructures (CFN),\\ Karlsruhe Institute of Technology, Karlsruhe, Germany\\
$^3$Theory and Simulation of Materials, Instituto de Ciencia de Materiales de Madrid, CSIC, Cantoblanco 28049, Madrid, Spain\\
$^4$Instituto de F\'{\i}sica Fundamental, CSIC, Serrano 113-bis, 28006 Madrid, Spain\\
$^5$Departamento de Qu\'{\i}mica F\'{\i}sica, Universidad del Pa\'{\i}s Vasco - Euskal Herriko Unibertsitatea, Apdo.\ 644, 48080 Bilbao, Spain\\
$^6$IKERBASQUE, Basque Foundation for Science, Alameda Urquijo 36, 48011 Bilbao, Spain}\\
(Dated: December 1, 2010)\\}
\end{center}


\twocolumngrid
The system of a flux qubit coupled to an $LC$ resonator can be modeled using the Hamiltonian
\begin{multline}
\label{eq1a}
H_E=\frac{\hbar\omega_q}{2}\sigma_z+\hbar\omega_r\left(a^{\dag}a+\frac{1}{2}\right)\\+\hbar g\left(\cos(\theta)\sigma_z-\sin(\theta)\sigma_x\right)(a+a^{\dag}),
\end{multline}
with $\hbar\omega_q\equiv\sqrt{\epsilon^2+\Delta^2}$, $\epsilon=2I_p(\Phi-\Phi_0/2)$ and $\tan(\theta)\equiv\Delta/\epsilon$. If the rotating-wave approximation is applied, Eq.~\ref{eq1a} becomes
\begin{multline}
\label{eq2a}
H_{JC}=\frac{\hbar\omega_q}{2}\sigma_z+\hbar\omega_r\left(a^{\dag}a+\frac{1}{2}\right)\\-\hbar g\sin(\theta)\left(\sigma_+a+\sigma_-a^{\dag}\right),
\end{multline}
known as the Jaynes-Cummings [JC] model.

A least-squares fit of the full spectrum of the system using Eq.~\ref{eq1a} can be seen in Fig.~\ref{eq1a} (solid black line), with fitted parameters $g/2\pi=0.82\pm0.03$~GHz, $\Delta/h=4.20\pm0.02$~GHz, $I_p=500\pm10$~nA, $\omega_r/2\pi=8.13\pm0.01$~GHz. Eq.~\ref{eq2a} is also plotted using these fitted parameters (dashed blue line). No significant difference can be observed between the two curves, except a small deviation at the symmetry point ($\Phi=\Phi_0/2$) for all transitions.

\begin{figure}[!htb]
\begin{center}
\includegraphics[width=\columnwidth]{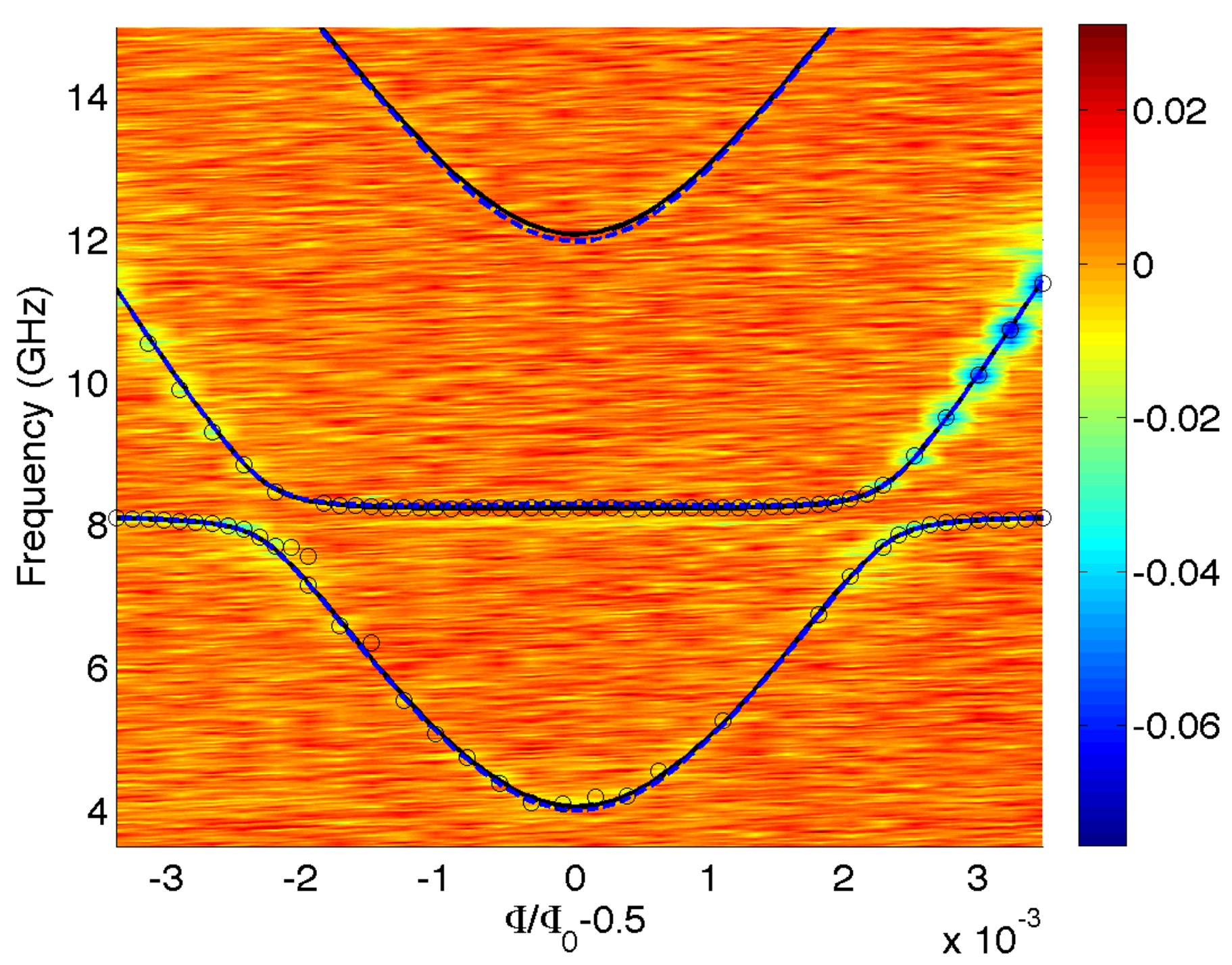}
\caption{Spectrum fitted using Eq.~\ref{eq1a} (solid black line). In dashed blue is a plot of Eq.~\ref{eq2a} (the JC model) using the fitted parameters.}
{\label{fig1a}}
\end{center}
\end{figure}

\begin{figure}[!htb]
\begin{center}
\includegraphics[width=\columnwidth]{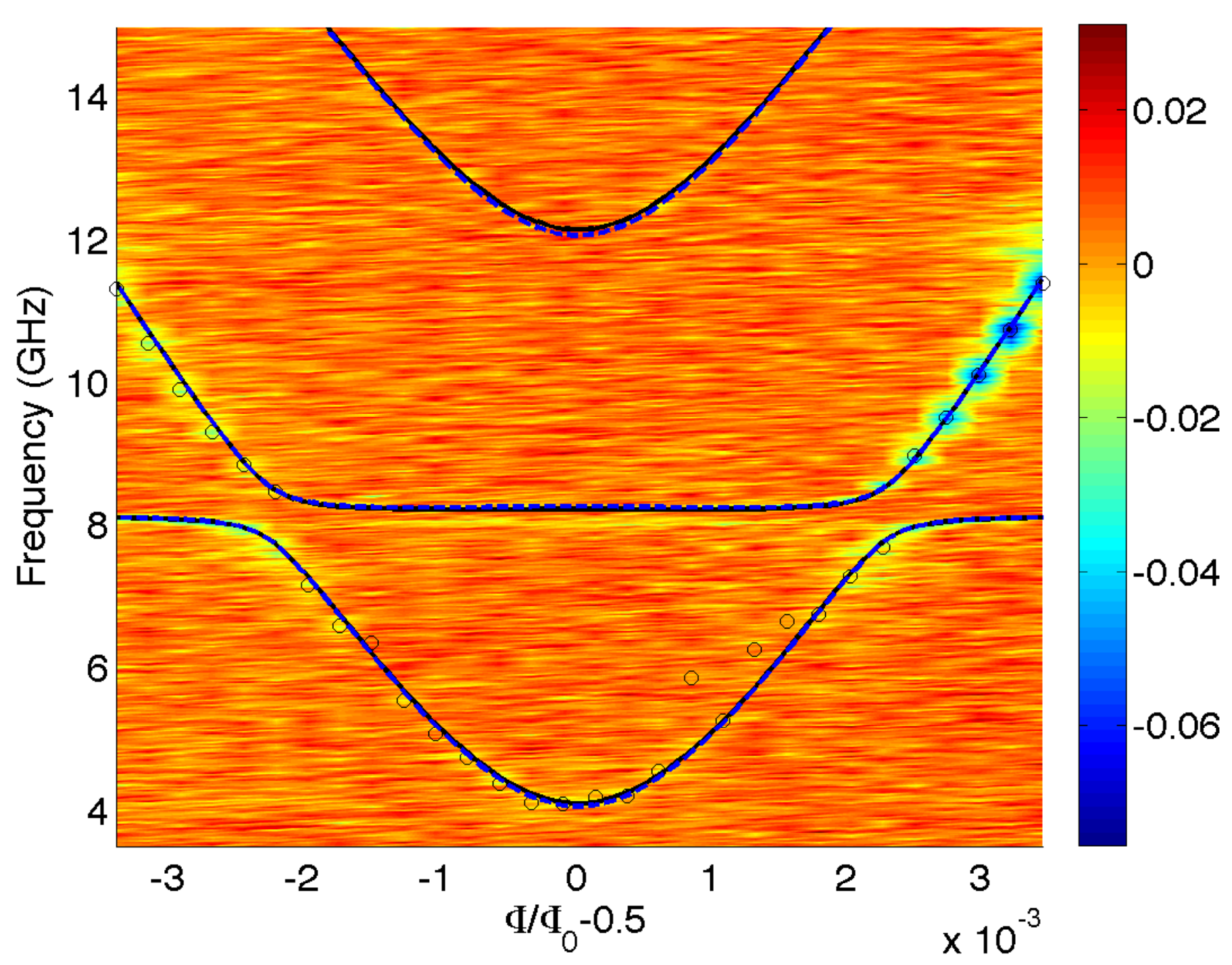}
\caption{Spectrum fitted using Eq.~\ref{eq2a}, the JC model (dashed blue line). In solid black is a plot of Eq.~\ref{eq1a} using the fitted parameters.}
\label{fig2a}
\end{center}
\end{figure}

A fit of the same spectrum using the JC model (Eq.~\ref{eq2a}) can be performed. This can be seen in Fig.~\ref{fig2a} (dashed blue line), with fitted parameters $g/2\pi=0.72\pm0.02$~GHz, $\Delta/h=4.21\pm0.02$~GHz, $I_p=500\pm10$~nA, $\omega_r/2\pi=8.13\pm0.01$~GHz. Eq.~\ref{eq1a} is plotted using these fitted parameters (solid black line). The difference between the two curves is similar to Fig.~\ref{fig1a}, with a small deviation at the symmetry point of the qubit.

\begin{figure}[!htb]
\begin{center}
\includegraphics[width=\columnwidth]{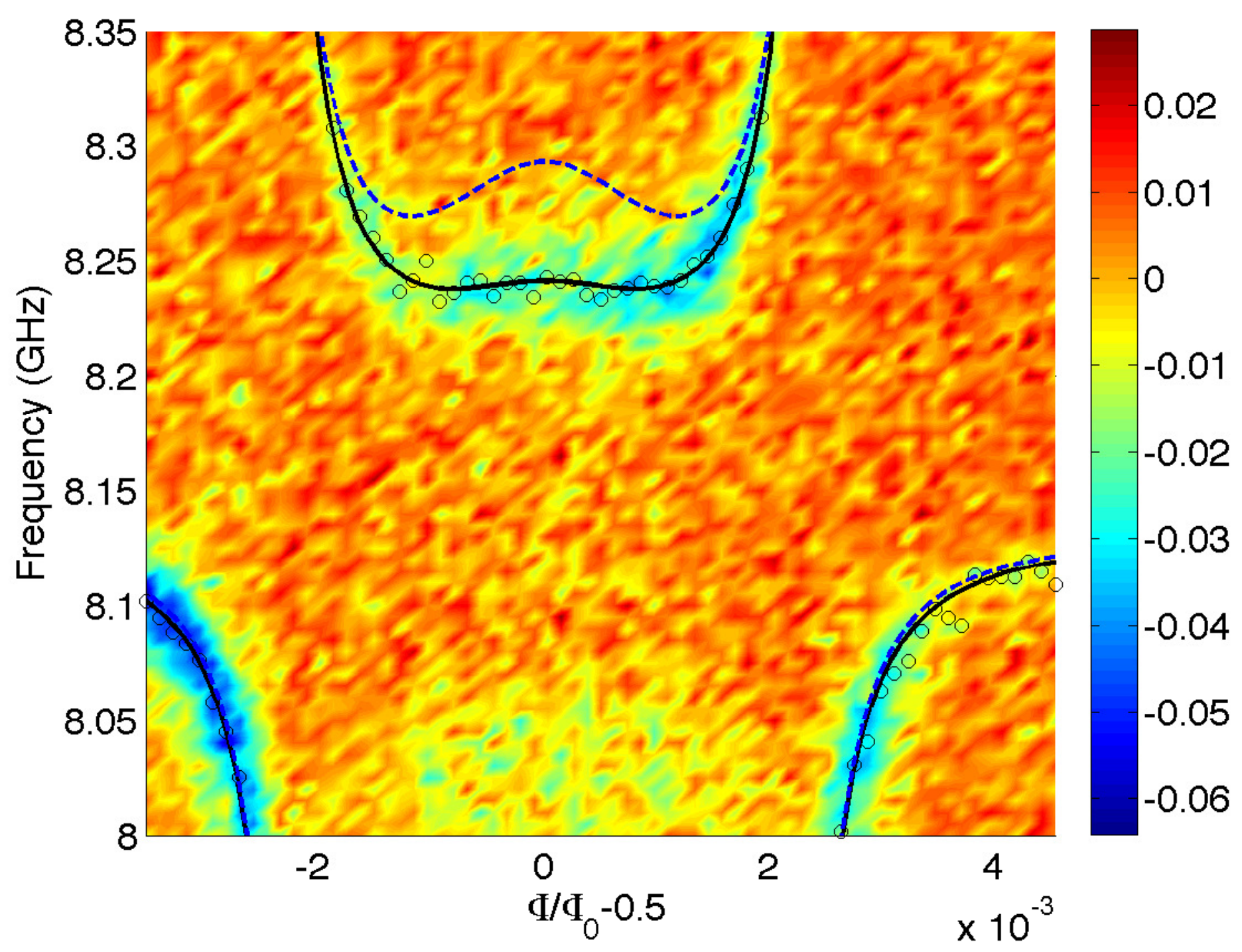}
\caption{Zoom in of Fig.~\ref{fig1a} around 8~GHz with the spectrum fitted using Eq.~\ref{eq1a} (solid black line). In dashed blue is Eq.~\ref{eq2a} using the fitted parameters.}
{\label{fig3a}}
\end{center}
\end{figure}

To observe more clearly the deviations between the exact model and te JC model, a zoom in is made of the spectrum at the region near 8 GHz. Fig.~\ref{fig3a} is a zoom in of Fig.~\ref{fig1a} showing the fit to Eq.~\ref{eq1a} (solid black line). Also the JC solution of Eq.~\ref{eq2a} is plotted (dashed blue line) using the same fitting parameters values to Eq.~\ref{eq1a}. Equation~\ref{eq1a} fits the data in all points (open circles represent Lorentzian fits to each data trace), while Eq.~\ref{eq2a} deviates at the symmetry point of the qubit. The deviation is attributed to the counter-rotating terms that were neglected by applying the rotating-wave approximation in Eq.~\ref{eq1a}.

\begin{figure}[!htb]
\begin{center}
\includegraphics[width=\columnwidth]{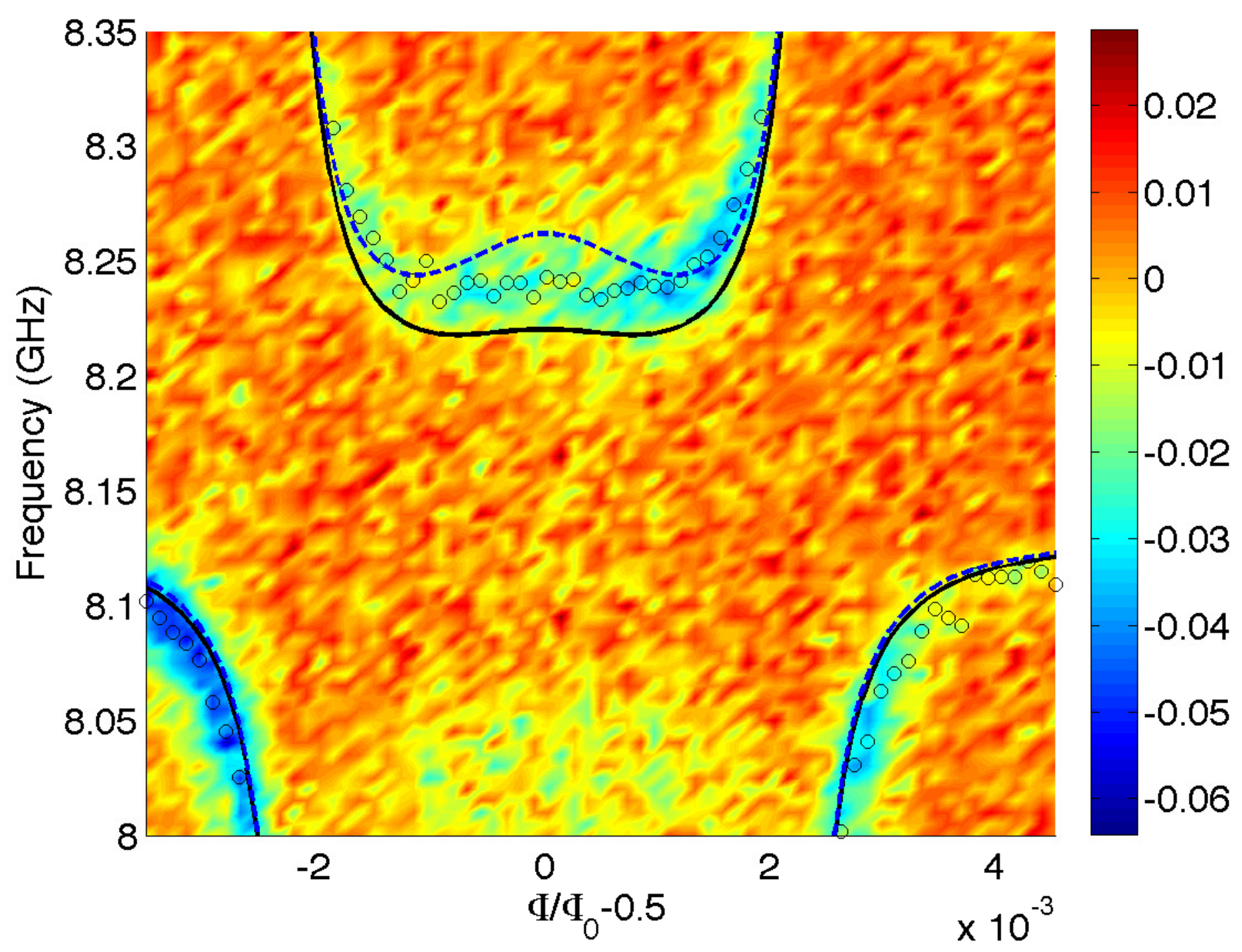}
\caption{Zoom in of Fig.~\ref{fig2a} around 8~GHz with the spectrum fitted using Eq.~\ref{eq2a}, the JC model (dashed blue line). In solid blue is Eq.~\ref{eq1a} using the fitted parameters.}
\label{fig4a}
\end{center}
\end{figure}

Figure \ref{fig4a} shows a zoom in of Fig.~\ref{fig2a} with the dashed line representing the fit to the JC solution Eq.~\ref{eq2a}. Also Eq.~\ref{eq1a} is plotted (solid black line) using the fit parameter values of Fig.~\ref{fig2a}.

In this case the best fit of the JC model (dashed line) does not fit all data points, in particular it does not fit the ones around $\Phi=\Phi_0/2$. On the other hand, Eq.~\ref{eq1a} using the fitted parameters from Fig.~\ref{fig2a} leads to lower values of the transition near 8.22~GHz than Fig.~\ref{fig3a}.

If the JC model was valid the fits in Fig.~\ref{fig1a} (and Fig.~\ref{fig3a}) and \ref{fig2a} (and Fig.~\ref{fig4a}) should lead to the same result. This is not the case, indicating that the rotating-wave approximation is not applicable. This is most clearly seen in Fig.~\ref{fig4a}, where the Jaynes-Cummings model fails to fit all data points, in particular in the range where the counter-rotating terms included in Eq.~\ref{eq1a} have their largest contribution providing maximum Bloch-Siegert shift.

\end{document}